\documentclass[letterpaper,twocolumn,10pt]{article}
\usepackage{usenix-2020-09}

\usepackage{multirow}

\usepackage{tikz}
\usepackage{amsmath}

\usepackage{amssymb}
\usepackage{listings}
\usepackage{xspace}

\newcommand\sname{BRF\xspace}
\newcommand\Sname{BRF\xspace}

\newcommand\CVEone{CVE-2022-2905\xspace}
\newcommand\CVEtwo{Vulnerability 2\xspace}
\newcommand\CVEthree{CVE-2023-0160\xspace}

\begin{document}

\title{BRF: eBPF Runtime Fuzzer}

\author{
{\rm Hsin-Wei Hung}\\
University of California, Irvine
\and
{\rm Ardalan Amiri Sani}\\
University of California, Irvine
}

\maketitle

\begin{abstract}

The eBPF technology in the Linux kernel has been widely adopted for different applications, such as networking, tracing, and security, thanks to the programmability it provides.
By allowing user-supplied eBPF programs to be executed directly in the kernel, it greatly increases the flexibility and efficiency of deploying customized logic.
However, eBPF also introduces a new and wide attack surface: malicious eBPF programs may try to exploit the vulnerabilities in the eBPF subsystem in the kernel.

Fuzzing is a promising technique to find such vulnerabilities.
Unfortunately, our experiments with the state-of-the-art kernel fuzzer, Syzkaller, shows that it cannot effectively fuzz the \textit{eBPF runtime}, those components that are in charge of executing an eBPF program, for two reasons.
First, the eBPF verifier (which is tasked with verifying the safety of eBPF programs) rejects many fuzzing inputs because (1) they do not comply with its required semantics or (2) they miss some dependencies, i.e., other syscalls that need to be issued before the program is loaded.
Second, Syzkaller fails to attach and trigger the execution of eBPF programs most of the times.

This paper introduces the BPF Runtime Fuzzer (\sname), a fuzzer that can satisfy the semantics and dependencies required by the verifier and the eBPF subsystem.
Our experiments show, in 48-hour fuzzing sessions, \sname can successfully execute 8$\times$ more eBPF programs compared to Syzkaller.
Moreover, eBPF programs generated by \sname are much more expressive than Syzkaller's.
As a result, \sname achieves 101\% higher code coverage.
Finally, \sname has so far managed to find 4 vulnerabilities (some of them have been assigned CVE numbers) in the eBPF runtime, proving its effectiveness.

\end{abstract}

\section{Introduction} 
\label{sec:intro}

Extended Berkeley Packet Filter (eBPF) is a rapidly evolving technology in the Linux kernel that enables generic programmability in the kernel space.
Originally, the Classic Berkeley Packet Filter (cBPF) was developed specifically for filtering network packets.
It allowed user-supplied programs to be loaded and executed in the kernel space to inspect packets and decide whether to allow or reject them.
A virtual machine in the kernel interpreted the simple cBPF bytecodes of the program.
In 2014, with new instructions and an enhanced virtual machine, eBPF was introduced~\cite{ebpfintrocommit}. 
As a result of the greater generic programmability and persistent data storage (i.e., maps)~\cite{ebpf1}, it has quickly gained traction in different domains in the kernel.
Not only is it widely adopted in different places in the networking stack, it is also integrated with kernel tracing, debugging, and auditing frameworks.

While eBPF brings extensibility and performance to the Linux kernel, it also introduces risks as user-supplied programs run in the kernel.
To ensure that an eBPF program, potentially supplied by attackers, can run safely in the kernel space, the eBPF verifier checks the program before loading it to make sure it will not execute indefinitely or access invalid memory.
Obviously, the correctness of the verifier is critical to the safety of kernel.
Therefore, many recent works~\cite{gershuni2019simple, wang2014jitk, nelson2020specification, bhat2022formal, van2020synthesizing, vishwanathan2021semantics, BPFFuzzer, nilsen2020fuzzing, BPFFuzzer2, scannell2020ebpffuzzer} have proposed different ways to test or verify it.

However, we note that eBPF has a significant footprint in the kernel beyond the verifier, i.e., the eBPF runtime, which executes an eBPF program after it passes the verifier.
The runtime components mainly consist of the execution environment (i.e., the interpreter or the just-in-time (JIT) compiler) and functionality that cannot be realized using only eBPF bytecodes (e.g., helper functions and maps).
Therefore, it is critical to find and fix the vulnerabilities in the eBPF runtime to prevent exploits.

Fuzzing is a promising technique to find such vulnerabilities.
Unfortunately, our experiments with the state-of-the-art kernel fuzzer, Syzkaller, shows that it cannot effectively fuzz the eBPF runtime for two reasons.
First, the aforementioned eBPF verifier rejects many fuzzing inputs because (1) they do not comply with its required semantics or (2) they miss some dependencies, i.e., other syscalls that need to be issued before the program is loaded.
Second, Syzkaller fails to attach and trigger the execution of eBPF programs most of the times.
We briefly discuss these issues here.

\noindent\textbf{Program semantics.}~~
Similar to other language processors, inputs to the eBPF subsystem (i.e., eBPF programs) pass through both syntax and semantic checks, and are then converted into low-level machine code (if JIT is enabled).
More specifically, input eBPF bytecode first go through a series of checks in the verifier.
Not only does the bytecode have to conform to the eBPF instruction set format (i.e., syntax), the control and data flow also need to conform to requirements imposed by the verifier to safeguard the kernel (semantics).

\noindent\textbf{Program dependencies.}~~
In eBPF, a series of preparatory syscalls are often needed to be made in order to correctly load the bytecode into the kernel.
The sequence and arguments of these syscalls may depend on the eBPF program.
For instance, loading a program that uses maps will require making syscalls to create compatible eBPF maps in the kernel in the first place.
The program also needs to be rewritten to refer to the external resources such as the just created maps, which is referred as \textit{relocation} in eBPF.

\noindent\textbf{Program execution.}~~
Finally, loaded eBPF programs need additional syscalls to be executed.
Since eBPF programs are triggered by kernel events, they first need to be correctly attached to the entry points.
Then, corresponding events need to created to trigger the execution of the programs.

To demonstrate Syzkaller's ineffectiveness in fuzzing the eBPF runtime, we perform an experiment.
During 48-hour fuzzing sessions, only 19.5\% of \texttt{BPF\_PROG\_LOAD} syscalls succeed in passing the verifier.
And the programs that pass the verifier are simple.
For example, the average number of instructions in a program is 4.86, showing that 50\% of the successfully loaded programs effectively contain less than 4 intructions to fuzz the runtime (The last instruction must be a \texttt{BPF\_EXIT}). 
Moreover, only 3,648 programs are actually executed after being loaded!

In this work, we set out to tackle these challenges to effectively fuzz the eBPF runtime.
Our solution is \sname\footnote{We will open source \sname.}, a fuzzer which is able to generate fuzzing inputs that, on the one hand, satisfy both the eBPF semantic constraints and eBPF program dependencies, and on the other hand, are expressive enough to explore different execution paths within the runtime.
Moreover, \sname attempts to attach and execute the successfully loaded programs.

We address three important challenges in \sname.
First, \sname produces semantic-correct eBPF programs efficiently
by using source-code level input generation/mutation and built-in semantic rules.
By generating/mutating eBPF programs at the source-code level and compiling them into actual fuzzing inputs, the compiler will automatically take care of the basic semantics of a program.
For example, a branch instruction will not jump to invalid locations, or an instruction will not access invalid stack memory unless we explicitly perform pointer arithmetics.
To further comply with additional semantics imposed by the eBPF verifier, we study the verifier, extract the rules and integrate them with the program generation/mutation logic.
Second, to tackle the challenge of syscall dependencies to eBPF programs, we study the relationship between the preparatory syscalls and eBPF programs, and then, in addition to random syscalls, we also generate preparatory syscalls with constrained arguments.
Finally, we generate syscalls to attach and trigger the eBPF programs.

Using extensive experiments, we show that 98\% of the fuzzing inputs generated by \sname succeed in passing the verifier.
Moreover, a large percentage of these programs are successfully attached to corresponding entry points and subsequently executed.
Overall, in 48-hour fuzzing sessions, \sname manages to execute 8$\times$ more eBPF programs than Syzkaller.
Furthermore, the programs successfully loaded by \sname are more expressive compared to programs successfully loaded by Syzkaller, i.e., they include 3.4$\times$ more instructions, 27.4$\times$ more calls to helper functions, and 17.1$\times$ more use of maps.
As a result, \sname can cover 101\% more basic blocks in the eBPF runtime when compared with Syzkaller.
Finally, \sname has so far managed to find 4 new vulnerabilities (some of which are assigned CVE numbers), proving its capability in finding vulnerabilities in the heavily-shielded runtime components.

\section{Background}
\label{background}

\subsection{Workflow of eBPF}
We have witnessed a wide and rapid adoption of eBPF in areas such networking, tracing, and security~\cite{ebpfsummit}.
For different use cases, there exist corresponding program types (e.g., \texttt{BPF\_PROG\_TYPE\_SOCKET\_FILTER} for filtering packets and \texttt{BPF\_PROG\_TYPE\_LIRC\_MODE2} for decoding infrared signals).
The program types limit the resources they can access (e.g., \texttt{BPF\_PROG\_TYPE\_LIRC\_MODE2} should not be able to access network packets).
We further breakdown the workflow into three phases: loading, attaching, and execution.

\noindent{\textbf{Loading.}}
An eBPF program that a user wants to execute first needs to be loaded into the kernel.
This involves loading BPF type format (BTF) information, creating eBPF maps, relocating the eBPF program and finally loading the program.
First, since an eBPF program may point to some variables in the kernel, BTF information is needed to resolve the references.
Next, eBPF maps used in the eBPF program need to be created by calling BPF syscalls.
Then, an eBPF program with instructions referring to external resources (i.e., kernel variables, maps) needs to be rewritten to point to the actual resources, which is also known as relocation.
Finally,
the eBPF verifier checks the program for the safety.
After that, it will be compiled by the JIT compiler if enabled. 
The kernel will return a file descriptor of the loaded program on success.

\noindent{\textbf{Attaching.}}
Once the eBPF program is loaded, the user can attach it using the aforementioned file descriptor to hooks in the kernel, which will be the entry points of the program.
The hooks can be functions in the network stacks, kernel functions, security hook, etc.

\noindent{\textbf{Execution.}}
Once the aforementioned hooks are triggered in the kernel, the attached program is executed either by an interpreter or natively if the in-kernel eBPF just-in-time (JIT) compiler is enabled.
Depending on the hook, different data will be passed to a program as the argument, which is called \textit{context}.
During the execution, the eBPF program can store or read data in eBPF maps.
It can also interact with the kernel through a predefined set of \textit{helper functions} in the kernel,
which allow a program to, for example, access maps, retrieve kernel information, or print messages.
Finally, an eBPF program may return an integer value, which will be interpreted by the kernel according to the program type.
For example, a socket filter program may return 0 to instruct the kernel to drop the packet.

\subsection{eBPF Verifier}

To ensure programs supplied by user space programs can be safely executed in the kernel, eBPF verifier statically checks them during loading.
Specifically, it mainly prevents unbounded execution, invalid jump, invalid memory access, and leak of sensitive kernel data.
To do so, the verifier first checks if a program contains loops in the control flow.
Then, it walks through the instructions and performs checks specific to the type of the instruction.
As it traverses the program, the verifier keeps track of the register state, which includes the potential ranges and types of values in the registers.
Some examples of the value types are normal scalar values (\texttt{SCALAR\_VALUE}), pointers to program context (\texttt{PTR\_TO\_CTX}), pointers to map (\texttt{CONST\_PTR\_TO\_MAP}), and pointers to stack memory (\texttt{PTR\_TO\_STACK}).
As a result, the verifier is able to determine whether a pointer dereference is safe.
For example, an instruction will only access the valid fields in the context with correct permission, or only pointers to valid stack memory can be dereferenced.
In addition, since the verifier can track the propagation of pointer values, leaking them to the user space (e.g., directly or indirectly through helper functions and return values) can be prevented.
We will discuss the verifier rules in more depth in \S\ref{sec:semantic}.

\subsection{BPF Runtime}

We define the eBPF runtime components as the parts of the eBPF subsystem that are executed once a eBPF program pass the safety checks from the verifier.
We identify four major components: JIT compiler, interpreter, eBPF maps, and helper functions.

\noindent{\textbf{JIT compiler.}}
In systems running on supported architectures (e.g., x86 and ARM), verified eBPF programs can be further compiled into native machine code to speed up the performance by the JIT compiler built into the kernel.

\noindent{\textbf{Interpreter.}}
If JIT is disabled or not supported by the architecture, the eBPF interpreter will be in charge of execution of programs, i.e., decoding eBPF bytecodes on the fly and executing them.

\noindent{\textbf{eBPF maps.}}
One significant improvement of eBPF over cBPF is persistent storage that preserves data even after programs terminate.
There are 29 different types of eBPF maps provided in Linux v5.15.
In general, they are key-value stores but differ in the underlying data structure (e.g., hash table, array, or ring buffer) or the type of elements being stored, e.g., generic data type, socket, or control group (cgroup).
In addition to eBPF programs, maps can also be accessed by user space programs using BPF syscalls.

\noindent{\textbf{Helper functions.}}
eBPF programs can interact with the kernel through helper functions, which are a predefined set of functions hard-coded in the kernel.
There are 175 helper functions in Linux v5.15.
A major portion of the helpers are used for accessing or manipulating eBPF maps.
For example, \texttt{bpf\_map\_lookup\_elem, bpf\_map\_update\_elem,} and \texttt{bpf\_map\_delete\_elem} are used for retrieving, modifying, and deleting an element in a map.
Some examples for other helper functions include printing debug messages, getting the \texttt{task\_struct} of the current task, and redirecting a packet to another network device.
Note that the availability of helper functions depends on the program type
as their usages only make sense under certain scenarios and privileges.
Also, unlike eBPF maps, helper functions can only be invoked by eBPF programs.

We design \sname to fuzz all these runtime components.
We specifically think that having the ability to fuzz eBPF maps and helper functions are important since as the eBPF technology finds new applications in the kernel, the number of eBPF maps and helper functions will for sure continue to grow.

\section{Overview}
\label{sec:design}

\subsection{Goals}

In this work, we aim to fuzz the eBPF runtime components.
The runtime primarily consists of the JIT compiler, the interpreter, the eBPF maps, and helper functions.
To achieve this, we have three goals.

\noindent{\textbf{Goal I: Generating semantic-correct eBPF programs.}}
Since most of the runtime components in the kernel space are only accessible through eBPF programs (except maps) instead of BPF syscalls from user space programs,
we need to generate eBPF programs and let them test these components.
Because the eBPF verifier enforces additional semantics in addition to C semantics on the eBPF programs for safety reasons and reject incorrect programs,
the randomly generated eBPF programs need to be semantic-correct in order to pass the heavy scrutiny of the verifier.

\noindent{\textbf{Goal II: Generating fuzzing inputs that meet syscall dependencies.}}
A fuzzing input generated by a fuzzer is a program consists of syscalls.
For the randomly generated eBPF programs, to pass the verifier to be loaded into the kernel, a series of syscalls need to be made.
The sequence and arguments of syscalls depend on the eBPF program.
Therefore, to effectively test the runtime, a fuzzing input should at least contain these syscalls that satisfy the eBPF program dependencies.

\noindent{\textbf{Goal III: Generating syscalls that attach and trigger eBPF programs.}}
After eBPF programs pass the verifier, to execute them, they first need to be attached to the hooks.
Then, events corresponding to the hooks need to generated in order to trigger them.
Thus, a fuzzing input should also include syscalls that can attach eBPF programs and trigger the execution.

\begin{figure}
\centering
\includegraphics[width=1.0\columnwidth]{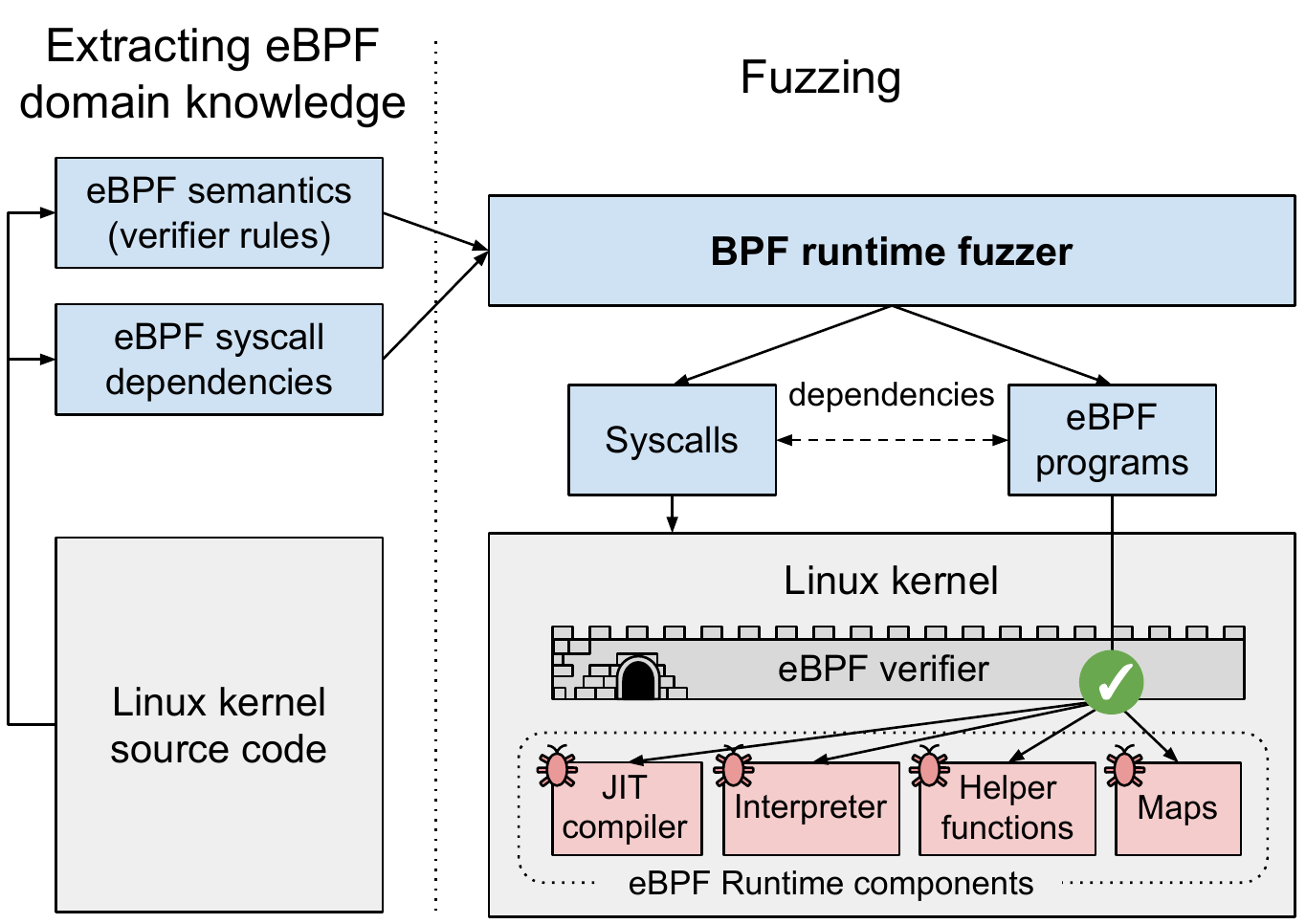}
\vspace{-0.2in}
\caption{The design of \Sname}
\label{fig:design}
\vspace{-0.1in}
\end{figure}

\subsection{Design}

Figure~\ref{fig:design} shows the high-level idea of \sname.
We first extract the eBPF domain knowledge, which includes the eBPF semantics and the syscall dependencies, through manual study of the source code and scripts that automatically parse the source code.
Then, during the fuzzing phase, by leveraging the extracted domain knowledge, \sname is able to generate both semantic-correct eBPF programs.
It then generates the fuzzing input, which includes not only the eBPF program, but also the required syscalls to correctly load, attach, and execute the eBPF programs.
This way, \sname can reach deeply behind the eBPF verifier and effectively fuzz the BPF runtime.

\begin{figure}
\centering
\includegraphics[width=1.0\columnwidth]{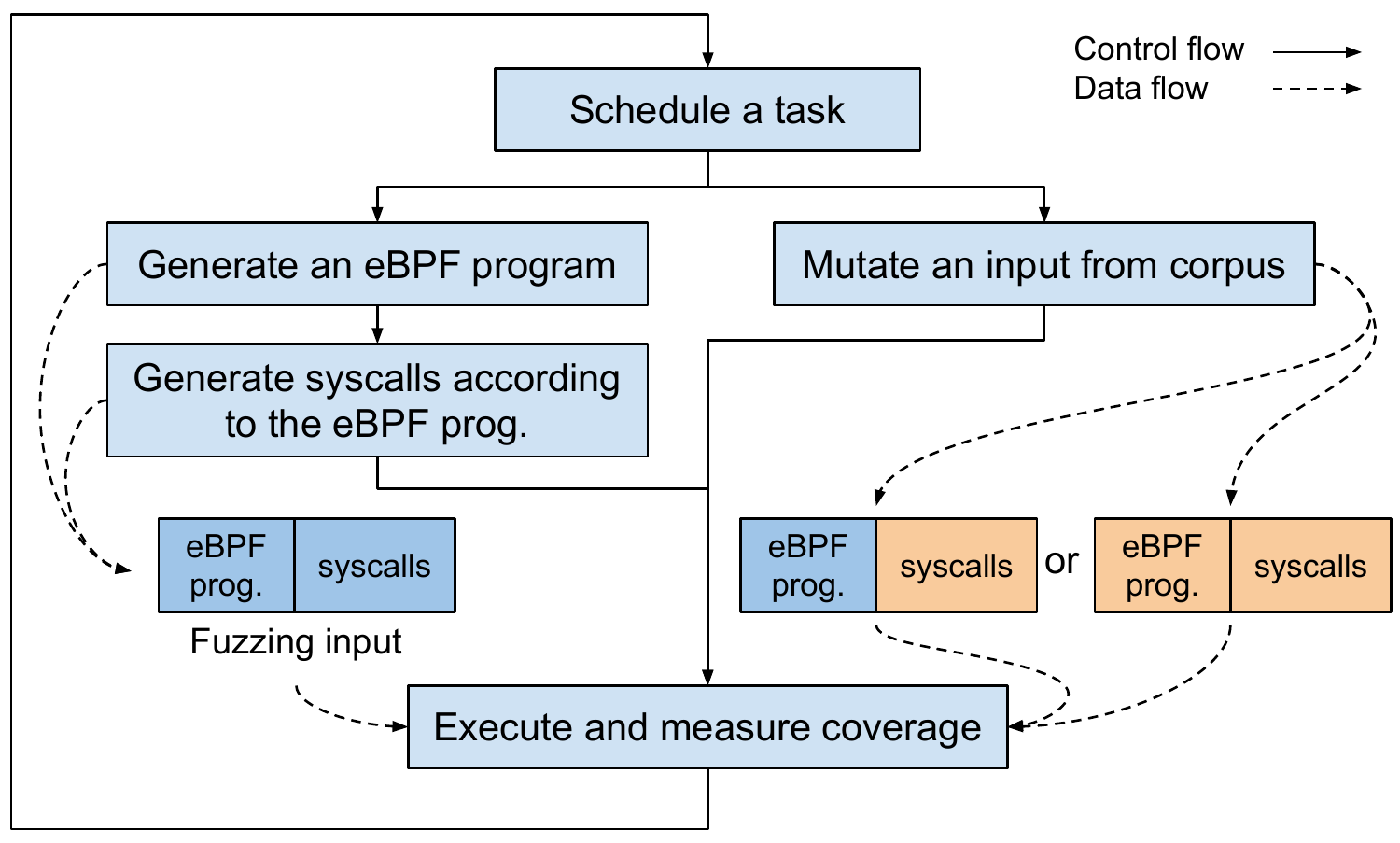}
\vspace{-0.1in}
\caption{The workflow of \Sname, where the orange blocks denote the mutated inputs}
\label{fig:workflow}
\vspace{0.0in}
\end{figure}

\subsection{Workflow}
\Sname is a coverage guided fuzzer as shown in Figure~\ref{fig:workflow}.
In the main fuzzing loop, a fuzzing input, which consists of an eBPF program and a series of required syscalls, is generated.
This is done in two steps.
First, the scheduler may decide to generate a new eBPF program or mutate an eBPF program in the corpus.
Then, different syscalls are generated to form a fuzzing input.

In the first step, to generate a new semantic-correct eBPF program, \sname first randomly chooses a program type and a target helper function
since helper functions are one of the key runtime components that we want to test.
Then, it tries to generate the arguments of the helper function.
For each argument, it will generate the value based on the type.
In general, there are three ways to generate a specific type of argument:
First, some type of arguments can be generated by directly passing different values into the argument.
Examples are scalar values or pointers to stack memory.
Second, accessing different fields of the program context sometimes can result in different types of values.
For instance, a socket filter eBPF program has \texttt{struct sk\_buff} as the context.
Accessing the field \texttt{sk} in this structure will yield a \texttt{PTR\_TO\_SOCK\_COMMON} value.
Lastly, helper functions can return different types of values.
Therefore, \sname may generate other helper function calls and their arguments recursively.
Note that the generation logic are constrained based on the rules enforced by the verifier (\S\ref{sec:semantic}) so that the resulting eBPF program will not contain incorrect semantics.
The generated eBPF program source code will be compiled into eBPF bytecode and also serialized and stored into the corpus.

In the second step, \sname generates a user space program with different syscalls (i.e., fuzzing input).
First, syscalls necessary to load, attach and execute the eBPF program are generated.
The arguments of these syscalls are generated according to the eBPF program so that they are compatible (\S\ref{sec:dependency}).
Then, other random BPF syscalls are generated and appended to the fuzzing input as they can also access some runtime components.

After the eBPF program and the syscalls, serving as a fuzzing input, are generated, it will be executed.
During the execution, coverage information is gathered to guide the fuzzing.
That is, if a fuzzing input triggers new coverage, the input will be added to the corpus and mutated to facilitate exploring different execution paths.

If the scheduler chooses to mutate a fuzzing input from the corpus, it may decide to mutate the eBPF program or the syscalls.
To mutate an eBPF program, \sname first randomly selects an argument of a helper function call.
Then, the argument will be randomly generated again using the same logic for input generation.
The mutated source code will then be compiled to form the new mutated eBPF program.
Since maps may be inserted or removed from the mutated eBPF program during this process, the syscalls associated with the eBPF program also needs to be modified accordingly
so that it can still load the eBPF program correctly.
On the other hand, if the scheduler decides to only mutate the syscalls, only the randomly generated syscalls and their arguments can be mutated to ensure the eBPF program can still be loaded after the mutation.

\section{Generating Semantic-Correct \lowercase{e}BPF Programs}
\label{sec:semantic}

\begin{table} \renewcommand{\arraystretch}{1.1}
\centering
\fontsize{8}{9}\selectfont
	\begin{tabular}{|l|l|l|l|l|}
\hline
Type of error msg. & Syntax  & Semantic	& Verifier error & Other 	\\ \cline{1-5}
\# error msg. 	   & 34	     & 226	& 38		 & 20		\\ \cline{1-5}
\end{tabular}
\caption{\em Types of error messages in the eBPF verifier.
}
\label{table:rules}
\end{table}

To improve the effectiveness of the fuzzer in testing runtime components, generating semantic-correct eBPF programs is critical.
Although eBPF programs can be written in C and compiled into executable using LLVM, the verifier imposes many additional semantics to ensure they can be executed in the kernel space safely.
Not only are there many semantic checks, they are mostly \textit{path sensitive}, making the verifier logic very complex.
In the kernel source code, the main file that contains verifier logic exceeds 15,000 line of code, not to mention many of other program-specific or map-specific logic scattered elsewhere in the kernel.

\noindent\textbf{Key approach.}~~
Studying the verifier in order to identify all the semantic checks is a daunting tasks.
Fortunately, we have an observation that allowed us to systematically tackle this challenge.
More specifically, we noticed that whenever errors occur in the verifier, corresponding error messages will be printed.
As shown in Table~\ref{table:rules}, the error messages mainly include instructions with incorrect syntax, semantic violations, and verifier internal errors.
Therefore, our key approach is to use the error messages to enumerate all the semantic rules in the verifier.

Since some verifier rules can be easily satisfied even with random program generation, not all rules need to be taken into consideration.
We aim to only find the rules that prevent us from effectively fuzzing the runtime and then constrain our eBPF program generation logic according to these rules.
We first annotate the verifier error messages with line numbers and start running the fuzzer without implementing constraints.
That is, the fuzzer will randomly generate helper functions and their arguments.
Then, during the development of the fuzzer, we add a constraint associated with a verifier rule if the rules is triggered more than once per hour by the eBPF programs generated by the fuzzer.
Note that, syntax rules will be automatically satisfied by compiling the source code into eBPF bytecode.

By studying the 226 semantic rules, we are able to constrain the generation and mutation processes of the fuzzer.
We explicitly incorporate constrains for 82 semantic rules into the fuzzer; most of the rest will be implicitly taken care of by the compiler and the construct of the program we use.
We next discuss some important rules that we have dealt with.

\definecolor{dkgreen}{rgb}{0,0.6,0}
\definecolor{gray}{rgb}{0.4,0.4,0.4}
\definecolor{lightgray}{rgb}{0.9,0.9,0.9}
\definecolor{mauve}{rgb}{0.58,0,0.82}
\lstset{frame=tb,
  language=C,
  showstringspaces=false,
  columns=flexible,
  basicstyle={\scriptsize\bf\ttfamily},
  xleftmargin={0.6cm},
  numbers=left,
  keywordstyle=\color{blue}\bf\ttfamily, 
  commentstyle=\color{dkgreen},
  stringstyle=\color{mauve},
  breaklines=true,
  breakatwhitespace=true,
  tabsize=3,
  moredelim=**[is][\color{blue}]{@}{@}
}

\begin{figure}
\vspace{0.04in}
\scriptsize
\begin{lstlisting}
static const struct bpf_func_proto *
kprobe_prog_func_proto(enum bpf_func_id func_id, ...)
{
	switch (func_id) {
	case BPF_FUNC_perf_event_output:
		return &bpf_perf_event_output_proto;
	case BPF_FUNC_get_stackid:
		return &bpf_get_stackid_proto;
	...
	default:
		return bpf_tracing_func_proto(func_id, prog);
	}
}
\end{lstlisting}
\vspace{0.0in}
\caption{\em Simplified code showing how the verifier checks if a helper is available to kprobe eBPF programs}
\label{fig:get_func_proto}
\vspace{0.0in}
\end{figure}

\subsection{Helper Function Availability}

For different types of eBPF programs, the helper functions available to them are different.
This makes sense as different program types has different use cases and require different privileges.
For example, for a socket filer eBPF program that requires almost no special privilege to be loaded, it should not be able to read arbitrary kernel memory using \texttt{bpf\_probe\_read\_kernel}.
Therefore, when the verifier encounters an eBPF call instruction to a helper function, it calls \texttt{get\_func\_proto}, a member function of the program type, to retrieve the function pointer to the helper function.
An example is shown in Figure~\ref{fig:get_func_proto}, which is the \texttt{get\_func\_proto} of kprobe eBPF programs.
If the helper function is not available, the program will be rejected.

To deal with this, \sname first parses the source code of the Linux kernel and looks for the definition of \texttt{get\_func\_proto} for every program type.
Sometimes, the \texttt{get\_func\_proto} of a program type may recursively call the \texttt{get\_func\_proto} of another program type.
For example, \texttt{bpf\_tracing\_func\_proto} shown in Figure~\ref{fig:get_func_proto} is the \texttt{get\_func\_proto} of the tracing type eBPF programs.
Therefore, after \sname gathers all \texttt{get\_func\_proto} of every program type, those containing other \texttt{get\_func\_proto}s will need to be recursively substituted to produce the complete helper availability information.
Then, during fuzzing, when \sname wants to insert a helper function, it will only choose randomly from the ones available to the program types.

\begin{figure}
\vspace{0.04in}
\scriptsize
\begin{lstlisting}
static const struct bpf_reg_types
	scalar_types = { .types = { SCALAR_VALUE } };
...
static const struct bpf_reg_types *
	compatible_reg_types[__BPF_ARG_TYPE_MAX] = {
		[ARG_CONST_SIZE] = &scalar_types,
		...
}
\end{lstlisting}
\vspace{-0.1in}
\caption{\em Simplified code showing how the verifier checks if a helper is available to kprobe eBPF programs}
\label{fig:compatible_reg_types}
\vspace{-0.1in}
\end{figure}

\subsection{Helper Function Arguments}

eBPF verifier poses strict type checking on variables in the program.
One of such checks happens when passing them to helper functions.
As opposed to the C semantics, where one can pass almost any type of variables to function arguments and implicit type conversion will happen under the hood,
variables passed to arguments of helper functions have to be of compatible types.
For example, as Figure~\ref{fig:compatible_reg_types} shows, a constant size type argument only accepts scalar type variables.
This suggests that a variable of any pointer types can never be passed to the argument because it not only makes no logical sense, but also creates a path to leak kernel pointers to the user space.

To generate eBPF programs conforming to these semantics constraints, we first extract the compatibility information between variable types and argument types stored in the \texttt{compatible\_reg\_types}.
Besides, \sname automatically parses the kernel source code to extract helper function prototypes, which include the type of arguments.
Then, when generating an argument for a helper function, it randomly chooses a compatible type of variable according to the argument type declared in the prototype by directly creating one, calling another helper function, or accessing the program context.

\subsection{Variable Safety Checks}

To prevent unsafe memory accesses commonly seen in programs written in the C language, such as null pointer dereference or out-of-bound access, pointer type variables in eBPF needs to be checked before being dereferenced.
By comparing a pointer variable to be a known value or range, the verifier (which is tracking the possible values of variables storing in the register state) is able to determine whether a pointer dereference is safe.
Any unsafe pointer dereference will cause the verifier to reject the eBPF program.
The safety constraint not only applies to normal expressions but also to arguments of helper functions as pointers can be passed to them and then dereferenced in the kernel.

In detail, there are three different types of checks that are required to safely use different types of variables when passing them as helper function arguments.

\noindent\textbf{Pointer.}
In the eBPF verifier, there are types of pointers that point to external resources, such as map values, sockets, memory locations or buffers.
To make sure null pointer dereference will not happen in helper functions, these pointers need to be compared with \texttt{nullptr} before passing them as arguments.

\noindent\textbf{Size.}
When a helper function takes a pointer to an external memory region as an argument, the argument following the pointer will be the size which will be used in the helper function to access the memory.
The size needs to be compared with a constant value that is smaller than the valid range of the memory to make sure out-of-bound memory access will not happen.
Also, it might need to be compared with zero when the argument does not allow a zero size access.

\noindent\textbf{Packet.}
For a pointer to a external packet, it needs to be compared with pointers to the start and end of the packet to make sure the access is limited within the packet.

To satisfy these constraints, \sname keeps track of the valid values of pointers.
Then, the safety checks will be generated before using them as arguments in helper functions.

More specifically, when \sname tries to generate a helper function, it will first generate it arguments, which could come from the return value of another helper function, the program context or direct allocation.
Then, an \texttt{if} condition block wrapping the helper function call will be generated.
The predicates will be filled with the safety checks of the variables passing to the arguments \texttt{and}ed together, so that only when all arguments are safe to use, the helper function can be invoked. 

For example, when a pointer to a map value is returned from a helper function, \sname knows that the valid size of the pointer should be the size of the value of the map.
Therefore, when passing the pointer to another helper function that takes an pointer to a memory region argument and a size argument specifying the size of the memory that will be accessed in the helper function, two checks will be added.
First, \sname will add a check to make sure the pointer to the map value return from another helepr is not a \texttt{nullptr}.
Second, the size argument will be compared to the size of the map value, so that if another random value is passed to the size argument and is larger than the size of map value, the function will not be invoked.

Note that to prevent over-constraining (i.e., putting unnecessary checks on arguments), not all pointer arguments needs to be checked.
When tracking pointer values, \sname records whether a pointer can potentially be \texttt{nullptr} as some helper functions will always return non-null pointers according to the return value annotated in the prototype.
Then, \sname generates a \texttt{nullptr} check only if an pointer argument does not allow \texttt{nullptr} and the pointer can potentially be a \texttt{nullptr}.

\subsection{Program Context Access}

eBPF programs are invoked with program contexts as arguments.
Most of these are pointers to different structures depending of the type of the program and where they are attached.
Not all members within the structure may be read or written, and the accesses to the contexts will be checked by the verifier by calling \texttt{is\_valid\_access} to make sure they are not only within the structure, but also with the right permission.

We extract the permission and definition of contexts in different \texttt{is\_valid\_access}s, and use this information to generate random but valid context accesses.
When generating a variable for an argument by accessing contexts, \sname first determines if the argument is going to be read or written.
This depends on if the argument is of \texttt{ARG\_PTR\_TO\_UNINIT\_MAP\_VALUE} type, which suggests the pointer will be accessed in raw mode and could be written.
Then, only the fields with the correct permission are used for random selection.

Note that unlike other variables that are declared and assigned right before the helper, there should be only one variable that accesses a specific field.
Otherwise, the compiler optimization will introduce pointer arithmetic on the pointer to context, which violates another verifier rule.
Therefore, variables generated by \sname using context accesses are inserted at the beginning of the program and then reused when needed.

\begin{figure}
\vspace{0.04in}
\scriptsize
\begin{lstlisting}
  ...
  v4 = bpf_ringbuf_reserve(&map_1, v2, v3);
  if (v4) {
    v5 = bpf_ringbuf_query(&map_1, v4);
  }
  ...
  if (v4) {
    bpf_ringbuf_submit(v4, 0);
  }
  return 1;
}
\end{lstlisting}
\vspace{-0.1in}
\caption{\em An example of a semantic-correct eBPF program rejected by the verifier.}
\label{fig:ref_issue}
\vspace{-0.1in}
\end{figure}

\subsection{Reference}

eBPF programs have the ability to acquire references to some kernel resources through helper functions (e.g., acquiring a reference to a socket interface).
Therefore, it is important that when a program terminates, the reference to the resources are relinquished.

In \sname, in order to generate eBPF programs that satisfy the reference rules, a fix-up process is performed after an eBPF program is generated.
More specifically, \sname goes through the helper functions in a program.
If a reference acquired by a helper function never gets released by another helper function, a new reference-releasing helper function will be generated and inserted to the program.
On the other hand, when a reference-releasing helper function tries to release a variable that is not produced by a reference-acquiring helper function, \sname will add a helper function that returns a reference and then substitutes the original argument.

When developing \sname for this rule, we notice some false positives of the verifier.
That is, an eBPF program conforming to the safety requirement is rejected due to the limitation of the verifier implementation.
An example is shown in Figure~\ref{fig:ref_issue}.
In line 2, \texttt{bpf\_ringbuf\_reserve} will acquire a reference to an entry in a ring buffer, and therefore must be released before the program exit.
Although \texttt{bpf\_ringbuf\_submit} is called to release the reference in line 8 after checking if the reservation succeeds in line 7 (which is a necessary safety check),
the verifier will determine that there is reference leak if \texttt{v4} is null.
Therefore, instead of releasing the reference before the program exit, we generate the reference releasing function right after the use, which in this case is after line 4.

\section{\lowercase{e}BPF program dependencies}
\label{sec:dependency}

Being semantic-correct alone does not guarantee that an eBPF program will be successfully accepted by the verifier.
It is also necessary for the program to have the correct preparatory syscalls to be called beforehand in order to load it into the kernel.
These syscalls depend on the eBPF program to be loaded, which include BPF syscalls and other generic syscalls.
More specifically, in the loading process, these syscalls need to create compatible eBPF maps and relocate the eBPF program.
Here we describe these syscalls and their dependencies to the eBPF program.

\subsection{Creating Compatible Maps}

Among the preparatory syscalls, \texttt{BPF\_MAP\_CREATE} syscalls need to be first called to create eBPF maps referred by eBPF programs.
The arguments of the syscall decide the attributes (i.e., the type of the map, type of the key, type of the value, the maximum number of entries and the flag describing other properties) of the map to be created.
Creating a compatible map can be done in two steps: (1) selecting a compatible type of map and (2) generating valid attributes for the map.
The compatibility of the type of map depends on the type of the program and the helper function that takes the map as the argument.
The attributes of a map are constrained by the type of the map in addition to the program type and helper functions.
Failing to meet the constraints can result in the failure of map creation.
In some cases even if the map creation succeeds, the verifier later will reject the program during load time.
We further describe the constraints (i.e., the dependencies between programs and maps) and how \sname satisfies them.

\noindent{\textbf{Helper function.}}
During fuzzing input generation, an eBPF map is first generated when a helper function wants to use it as an argument.
A helper function may only be compatible with a certain type of maps.
An obvious example is that \texttt{bpf\_ringbuf\_output} will only accept \texttt{BPF\_MAP\_TYPE\_RINGBUF} type of maps.

Thus, when selecting the type of map to be generated, \sname randomly chooses a compatible one using the compatibility information extracted from the verifier in the function \texttt{check\_map\_func\_compatibility}.

\noindent{\textbf{Map type.}}
For different types of maps, the attributes have different constraints and will be checked during the creation.
For example, since a \texttt{BPF\_MAP\_TYPE\_CGROUP\_STORAGE} map only holds an entry local to a cgroup and is indexed by a 64-bit cgroup ID or \texttt{struct bpf\_cgroup\_storage\_key}, the key size can only be 64 bits or the size of a \texttt{struct bpf\_cgroup\_storage\_key}.
Besides, since the number of entries cannot be determined by user space programs, it should pass 0 as the number of max entries in the arguments.

Therefore, to make sure \texttt{BPF\_MAP\_CREATE} syscalls succeed, \sname generates the attributes according to the constraints of different map types.
There are four attributes; and the constraints we extract from the \texttt{map\_alloc} and \texttt{map\_alloc\_check} functions of different map types can be generally described as followed:
For the size of keys and the size of values, the constraints are the minimum size, maximum size and the alignment.
For the flags, valid flags for a specific map are first separated into groups, where only one flag can only be selected in each group.
Then, these selected flags are \texttt{or}-ed together to form the final flag value.
Finally, the constraint of the max entries is a single value that limits the upper bound of the random generated value.

\noindent{\textbf{Program type.}}
The type of a program may also affect the types of maps that can be used (i.e., certain maps are not compatible with certain program types).
For example, a tracing type program that will be attached to hooks that may sleep during execution can only use some basic hash and array maps.
The type of program may also affect the flags of the map.
For instance, the memory should be pre-allocated for perf event type eBPF programs, \texttt{BPF\_PROG\_TYPE\_PERF\_EVENT}.
Therefore, the flag, \texttt{BPF\_F\_NO\_PREALLOC}, shall not be used.

When generating a map, \sname first only selects from types of maps that can be used under the current program type, which is similar to how the compatibility of maps with helper functions is handled.
Then, attributes are generated based on the constraints coming from the map type.
For the attribute constraints introduced by the program types, \sname tries to fix the flags after all the attributes are generated depending on the programs type.

\subsection{Relocating eBPF programs}

eBPF programs need to be relocated when they call other eBPF programs or there are instructions that refer to maps, external symbols, or global variables.
In such case, a user space program that wants to load the program first needs to create the maps or programs or resolve the reference.
Then, the references in the instructions need to be updated.

\sname addresses the problem by inserting correct and immutable syscalls into the fuzzing inputs.
To generate the correct syscalls, we leverage an eBPF utility library, \texttt{libbpf}.
We directly invoke the API of \texttt{libbpf} that makes syscalls to load an eBPF program after parsing eBPF bytecode and other sections in the ELF file of the program.
Note that these preparatory syscalls in the fuzzing inputs will not be mutated.
As a result, instead of randomly generating these syscalls and hoping they are sufficient for loading an eBPF program, we can make sure an eBPF program will be loaded successfully first in each fuzzing input.

\section{Executing \lowercase{e}BPF programs}
\label{sec:dependency}

To execute the eBPF programs generated by \sname, two steps need to be done by the fuzzing inputs.
eBPF programs supplied by user space programs will not be automatically executed after being successfully loaded into the kernel.
They first need to be attached to hooks in the kernel.
Then, when the events corresponding to the hooks happen, they will be executed.

\subsection{Program attachment}

For some eBPF program types, the entry point information (i.e., where the program should hook to) is contained within the binary and therefore can be directly attached by calling \texttt{BPF\_PROG\_ATTACH} without specifying them in the arguments.
A customized section in the ELF binary of a program is used to specify the type of the program.
Sometimes, it may also contain the entry points.
For instance, an eBPF program with an ELF section named \texttt{kprobe/sys\_nanosleep} not only tells the user space programs loading it that it is a kprobe/eBPF program, it also says that the program should be attached to the entry point of the syscall, \texttt{nanosleep}.
In this case, \sname will use a fixed hook for every program types when generating the binary.

However, for some types of the programs, for flexibility reasons, the entry point information are not be specified in the binaries.
Instead, they are provided during attachment using the arguments of the syscall.
An example is \texttt{BPF\_PROG\_TYPE\_LIRC\_MODE2}, which enables decoding infrared signal using eBPF programs.
When attaching an LIRC eBPF program, an opened instance of the LIRC driver should be passed to the attach syscall to indicate from which devices the signal should be decoded by the program.
Therefore, to properly attach these types of eBPF programs, \sname further generates syscalls required to create and open the resources, and then use it for the attachment.

\subsection{Triggering the hooks}

After the eBPF programs are attached, besides letting them get triggered opportunistically by random events in the kernel, \sname further use two methods to increase the probability of their execution.

First, \sname generates syscalls to trigger 7 types of eBPF programs (4 tracing-related, 2 network-related, and \texttt{LIRC}).
For tracing type eBPF programs, we explicitly attach to events that happen frequently (e.g., tracepoints in the kernel scheduler, syscall entry points, or events signaled by hardware periodically).
For the two network-related programs, we call \texttt{recv} on the socket we created and write data to the socket as well to trigger the execution.
For \texttt{LIRC} programs, after attaching to the \texttt{LIRC} device, \sname generates syscalls that write signals to the device, so that the program will be invoked to decode the infrared signal.

Secondly, \sname leverages the BPF syscall \texttt{BPF\_PROG\_TEST\_RUN} to deal with eBPF programs associated with subsystems that are hard to set up and test due to their complexity.
Introduced for this exact same purpose, \texttt{BPF\_PROG\_TSET\_RUN} facilitates testing of a subset of program types by building a simulated environment in the kernel and then test-running the eBPF programs.
Therefore, \sname always generates a \texttt{BPF\_PROG\_TEST\_RUN} syscall after loading and attaching programs.

\section{Implementation}
\label{sec:implementation}

We implement \sname on top of Syzkaller in Go and C++ with 3000 LoC.
We extracted the constraints from Linux v5.15 by manually studying the verifier and partially parsing the source code automatically. 
Finally, to collect coverage of eBPF programs for guiding the fuzzing, we extend the coverage collecting framework in the Linux kernel, \texttt{kcov}, and instrument the kernel.

\subsection{Collecting Coverage of eBPF Programs}

While Syzkaller already collects coverage information of the fuzzer process to guide the fuzzing, it does not include the coverage of eBPF programs.
If not taken care of, eBPF programs that actually trigger new execution paths will not be prioritized, leading to sub-optimal fuzzing.

To collect coverage information of syscalls issued from the fuzzer process, Syzkaller leverages \texttt{kcov} in the Linux kernel.
Through compiler instrumentation, a \texttt{kcov} function will be invoked for every edge in the kernel.
For every thread that opens and enables \texttt{kcov}, the coverage information is then stored to a thread-specific memory region.
In other words, only coverage of syscalls from userspace processes are collected.

However, eBPF programs are triggered by different events depending on the program types and the hooks.
Therefore, they normally execute in contexts different from the original fuzzer process that loads and attaches the eBPF programs, for example, kernel threads or interrupt contexts.
As a result, even when a new helper function is invoked in an eBPF program, the eBPF program may not be prioritized since no positive feedback is generated.

Fortunately, Syzkaller supports collecting extra coverage information related to the fuzzer from other processes by utilizing the \textit{remote} coverage feature of \texttt{kcov}.
This would allow us to collect a kernel thread's coverage and then associate it with a user process.
It requires manually instrumenting the source code, which involves three steps.
First, the handle of the user space thread needs to be installed to the kernel thread of interest.
We do so when a user space program loads an eBPF program into the kernel space.
The handle will be stored into \texttt{struct bpf\_prog}.
Second, in the kernel thread, we need to annotate when to start recording the coverage by calling \texttt{kcov\_remote\_start} with a handle.
This is done at the beginning of the function \texttt{bpf\_prog\_run}, with the handle we store in the \texttt{struct bpf\_prog}.
Therefore, the coverage can be associated with the user space program regardless of where the eBPF program is executed.
Finally, we need to call \texttt{kcov\_remote\_stop} to stop recording coverage when the execution of the thread leaves the area of interest, which is the end of \texttt{bpf\_prog\_run}.

We further extend the \texttt{kcov} remote coverage API to make tracing eBPF program coverage possible.
Since the original remote coverage API allocates a large memory area for storing coverage in \texttt{kcov\_remote\_start} using \texttt{vmalloc}, 
it can only be invoked in context where sleep is permitted.
However, this is not the case for most of the eBPF program entry points.
Therefore, we create a preallocated version remote coverage API, which takes preallocated memory as argument when starting collecting coverage.
Then, for every eBPF program, we allocate a memory region during loading.

\section{Evaluation}
\label{sec:evaluation}

\def\mc#1#2{\multicolumn{#1}{c|}{#2}}
\def\mr#1#2{\multirow{#1}{*}{#2}}

\def\Tot{\mc{1}{Total}}
\def\Ts{\mc{1}{Total success}}
\def\Rs{\mc{1}{Success rate}}

\begin{table*}[ht] \renewcommand{\arraystretch}{1.1}
\centering
\fontsize{8}{9}\selectfont
	\begin{tabular}{|l|r|r|r|r|r|r|r|r|}
\hline
		& \mc{3}{Program loading syscall} & \mc{3}{Program attaching syscall} 	& \mc{2}{Program execution}	\\ \cline{2-9}
		& \Tot	    & \Ts 	& \Rs 	  & \Tot	& \Ts	   & \Rs	& \Ts*		& \mc{1}{Success rate**}	\\ \cline{1-9}
Syzkaller	& 176k $\pm$ 14k & 34k $\pm$ 7k		& 19.5\% $\pm$ 3.8\%  & 60k $\pm$ 17k & 16k $\pm$ 5k   & 26.5\% $\pm$ 0.7\%	& 12k $\pm$ 1k	& na 			\\ \cline{1-9}
\sname		& 176k $\pm$ 9k  & 172k $\pm$ 10k 	& 97.6\% $\pm$ 0.8\%  & 160k $\pm$ 8k & 145k $\pm$ 8k  & 89.9\% $\pm$ 1.2\%	& 97k $\pm$ 8k	& 66.8\% $\pm$ 3.7\%	\\ \cline{1-9}
\end{tabular}

\vspace{-0.05in}
\caption{\em Comparison of programs loaded, attached, and triggered by Syzkaller and \sname under 48-hour fuzzing sessions.
*Total success keeps track of whether any eBPF programs referenced in a fuzzing input are executed. Therefore it may contain duplicated eBPF programs.
**The success rate only counts the number of unique eBPF programs that are executed, which is not available in Syzkaller.
}
\vspace{-0.1in}
\label{table:fuzzing_stats}
\end{table*}

\begin{table} \renewcommand{\arraystretch}{1.1}
\centering
\fontsize{8}{9}\selectfont
	\begin{tabular}{|l|r|r|r|r|r|r|}
\hline
		& \mc{2}{\# Instructions}  	& \mc{2}{\# Helper functions}	& \mc{2}{\# Maps}		\\ \cline{2-7}
		& \mc{1}{Avg.}	& \mc{1}{Max}	& \mc{1}{Avg.}	& \mc{1}{Max}	& \mc{1}{Avg.}	& \mc{1}{Max}	\\ \cline{1-7}
Syzkaller	& 4.9  $\pm$ 0.1 & 16		& 0.4 $\pm$ 0.1	& 4	   	& 0.3 $\pm$ 0.0	& 4	\\ \cline{1-7}
\sname		& 16.4 $\pm$ 0.1 & 314		& 11.0 $\pm$ 1.2	& 66 	& 5.6 $\pm$ 0.7	& 18	\\ \cline{1-7}
\end{tabular}
\vspace{-0.00in}
\caption{\em Expressiveness of eBPF program generated and successfully loaded by Syzkaller and \sname.
}
\vspace{-0.15in}
\label{table:program_expressiveness_single}
\end{table}

\subsection{Fuzzing Effectiveness}

To evaluate the effectiveness of \sname in fuzzing the runtime components of the eBPF subsystem, we compare it with the state-of-art syscall fuzzer, Syzkaller.
In the evaluation, we assign the same amount of resources to the two fuzzers, which are five virtual machines, each with 4GB of RAM and eight fuzzer processes.
Besides, since we are only interested in fuzzing the eBPF subsystem, only BPF related syscalls are enabled.
To have a fair comparison, we enable all BPF-related syscalls for Syzkaller so that it is able to create the necessary resources for attaching programs and triggering the execution.
For example, \texttt{socket} and \texttt{setsockopt} are enabled so that if a \texttt{BPF\_PROG\_TYPE\_SOCKET\_FILTER} eBPF program is generated, there are syscalls to create a socket and attach the program to the socket.
For \sname, we only enable core eBPF syscalls as the fuzzer will generate the necessary syscalls.
In addition, since the JIT compiler is enabled by default on x86 for performance reasons, we report the fuzzing statistics under this configuration.
However, when running fuzzing experiments to find vulnerabilities, we try both kernels with and without JIT compiler so that we can also test the interpreter. 
We run the fuzzer on a machine with Intel Xeon E5-2697 v4 CPU for 48 hours and compare the results.

\noindent\textbf{Reaching the runtime components.}~~
First, we look at the three critical stages in the workflow of eBPF that will affect the fuzzing effectiveness: eBPF program loading, attaching and execution.
Since many of the runtime components, such as the interpreter, JIT compiler, and helper functions, can only be accessed by eBPF programs, to achieve high fuzzing effectiveness,
it is essential first for the programs to pass the verifier during the load time.
As shown in Table~\ref{table:fuzzing_stats}, 97.6\% of eBPF programs generated by \sname are able to pass the checks of the verifier,
showing that these fuzzing inputs are both semantic-correct and meet syscall dependencies.
Since we do not cover all semantic rules, there are still a small portion of fuzzing inputs that fail to pass the verifier.
On the other hand, with only syntax-awareness and little knowledge of syscall dependencies, only 19.5\% of programs generated by Syzkaller pass the verifier.

Here we report five most violated verifier rules (>100 violations in 48 hour) that bottleneck the fuzzing of the runtime for Syzkaller: 1) calling into invalid destination 2) not having \texttt{jmp} or \texttt{exit} as the last instruction 3) calling into a \texttt{btf\_id} which is not a kernel function 4) jumping out of range 5) calling a kernel function from non-GPL eBPF program.
Note that these bottlenecks are relatively simple.
The reason is that the semantic checks are layered and therefore the complex one are still masked by the simple ones.

After eBPF programs pass the verifier, they will be compiled by the JIT compiler if enabled, and the JIT compiler will be fuzzed.
Then, to exercise the interpreter (when JIT is disabled), helper functions and maps, programs first need to be attached.
For Syzkaller, due to the fact that only a few eBPF programs are loaded successfully, the program attaching syscall has few valid program \texttt{fd}s to start with.
Since Syzkaller only generates program attaching syscalls by chance, only 26.5\% of these syscalls succeed.
Note that, these successfully attached programs of Syzkaller could be the same eBPF programs.
Since Syzkaller generates syscalls randomly, it is possible that a fuzzing input retrieves a already loaded program pinned in BPF virtual file system.
While in \sname, we only count unique eBPF programs that are loaded.
In contrast, since \sname generates program attaching syscalls for every fuzzing input, 89.9\% of its attaching syscalls succeed.
Finally, \sname manages to trigger 66.8\% of the attached programs.
Whereas for Syzkaller, we are only able to collect the total number of eBPF programs executed with the possibility of duplication.
All and all, the advantages of \sname in loading, attaching, and triggering eBPF programs result in \sname executing 8$\times$ more programs than Syzkaller.

\noindent\textbf{Expressiveness of eBPF programs.}~~
Beside the success rate of loading, attaching, and executing eBPF programs, the expressiveness of the eBPF programs can also affect the fuzzing effectiveness.
Therefore, we quantify the expressiveness of the generated eBPF programs by measuring the average number of eBPF instructions, average number of calls to helper functions, and the average number of usage of maps within within successfully loaded eBPF programs.
Our results, shown in Table~\ref{table:program_expressiveness_single}, show that, on average, a program successfully loaded by \sname contains 3.4$\times$ more instructions, 27.4$\times$ more calls to helper functions, and 17.1$\times$ more use of maps, compared to a program successfully loaded by Syzkaller.
The largest program generated by \sname can contain 66 helper functions and 18 maps.
The higher effectiveness of \sname in both generating semantic-correct and expressive eBPF programs
help it reach broader and deeper in the eBPF runtime.

\begin{figure}
\centering
\includegraphics[width=1\columnwidth]{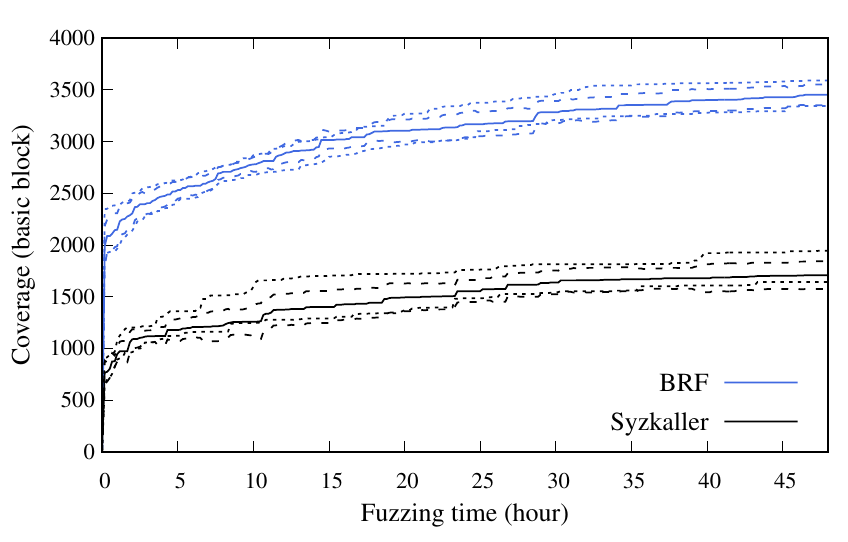}
\vspace{-0.35in}
\caption{\em Coverage of eBPF runtime components of \sname and Syzkaller over time. Solid line is median; dashed lines are confidence intervals; dotted lines are max/min }
\label{fig:coverage_core}
\vspace{-0.2in}
\end{figure}

\noindent\textbf{eBPF runtime coverage.}~~
We measure the ability of BRF and Syzkaller in covering different types of helper functions.
eBPF programs generated by \sname uses 155 different helper functions, while the ones generated by Syzkaller can only uses 124.

\noindent\textbf{Code coverage.}~~
To see how the aforementioned key metrics in the fuzzing effectiveness translate into the ability of exploring more execution paths, we record the coverage of Syzkaller and \sname in the eBPF runtime components.
We do so by only including the basic blocks in files that implement the eBPF runtime: JIT compiler, interpreter, helper functions and maps.
Therefore, the BPF syscalls, verifier and glue code that connects eBPF subsystem with other subsystems are not included.
The result is shown in Figure~\ref{fig:coverage_core}, where we find that during the 48-hour fuzzing sessions, \sname is able to continuously discover new paths.
In contrast, the coverage of Syzkaller plateau after 5 hours into the fuzzing.
At the end of the session, \sname is able to explore 101\% more basic blocks in the runtime components, showing its better ability in fuzzing eBPF runtime.

\subsection{Discovered Vulnerabilities}

Table~\ref{table:vulnerability} lists the vulnerabilities found by \sname.
Next, we discuss them in detail.

\begin{table*}[ht] \renewcommand{\arraystretch}{1.1}
\centering
\fontsize{8}{9}\selectfont
	\begin{tabular}{|l|c|c|c|c|c|c|l|}
\hline
		& \mr{2}{Verifier}  & \mc{4}{Runtime components}			& \mr{2}{New} & \mr{2}{Type of vulnerability}	\\ \cline{3-6}
		&		& JIT compiler	& Interpreter	& Helper functions	& Maps  &	     &						\\ \cline{1-8}
\CVEone	& \checkmark	& \checkmark	&		&		&     		& \checkmark & Out-of-bound access, information leakage	\\ \cline{1-8}
\CVEtwo	&		&		&		& \checkmark	&		& \checkmark & Deadlock, API misuse			\\ \cline{1-8}
\CVEthree	&		&		&		& \checkmark	& \checkmark	& \checkmark & Deadlock					\\ \cline{1-8}
Vulnerability 4	&		&		&		& \checkmark	& \checkmark	& \checkmark & Deadlock					\\ \cline{1-8}
\end{tabular}
\vspace{-0.0in}
\caption{\em Summary of vulnerabilities discovered by \sname.
}
\vspace{-0.1in}
\label{table:vulnerability}
\end{table*}

\begin{figure}
\vspace{0.04in}
\scriptsize
\begin{lstlisting}
DEFINE_BPF_MAP(map_0,
	BPF_MAP_TYPE_PROG_ARRAY, // type
	0,                       // map_flags
	uint32_t,                // key
	uint32_t,                // value
	36                       // max_entries
);
SEC("cgroup/sock_create")
int func(struct bpf_sock *ctx) {
	bpf_tail_call(ctx, &map_0, 49);
	return 0;
}
\end{lstlisting}
\vspace{-0.1in}
\caption{\em Simplified code illustrating a Proof-of-Concept (PoC) of \CVEone.}
\label{fig:cve_2022_2905}
\vspace{-0.1in}
\end{figure}

\noindent\textbf{\CVEone} is an out-of-bound heap memory access vulnerability.
As shown in Figure~\ref{fig:cve_2022_2905}, an attacker can probe into the kernel memory by calling \texttt{bpf\_tail\_call} in an eBPF program with a key larger than the \texttt{max\_entries} of the map.
\texttt{bpf\_tail\_call} is a helper function that allows the calling program to jump into another program stored in the map of the kind, \texttt{BPF\_MAP\_TYPE\_PROG\_ARRAY}.
When this malicious eBPF program is loaded into the kernel, it is able to pass the verifier due incorrect value range analysis in the verifier.
Therefore, when the x86 JIT compiler tries to compile the program, it will index the array of program stored in \texttt{bpf\_array->ptr} using the key.
This will cause an out-of-bound access in the heap area as only 36 entries is allocated to \texttt{bpf\_array->ptr} when creating the map prior to loading the program.

\Sname is able to discover the vulnerability because calling \texttt{bpf\_tail\_call} with a key larger than the \texttt{max\_entries} does not violate the semantic rules enforced by the verifier.
Instead, the helper function should perform the check on this argument during run time.
Since there is no input generation constraints on the key values, and it is possible for \sname to generate such input.
In fact, the verifier would have forced programs to be executed in interpreter mode if the key was not a known constant or was "out of the range" of the \texttt{max\_entries} instead of rejecting them.
However, since the value range analysis mechanism in the verifier over-approximates the range, a key larger than the \texttt{max\_entries} could be deemed within the range.
As a result, such kind of programs can be executed in JIT mode and trigger the vulnerability. 

\noindent\textbf{\CVEtwo}\footnote{We have reported this vulnerability to both the Linux kernel community and the vendor, and we believe a CVE number will be assigned in this process.} is a warning triggered when the helper function, \texttt{bpf\_probe\_read\_user}, is called in a non-maskable interrupt (NMI) context.
The warning comes from the API for copying data from user space memory, \texttt{copy\_from\_user\_nofault}, since it was intended to be called in user context.
The kernel developer of the eBPF subsystem, Alexei Starovoitov, further found two problems that came along with this mis-used API.
First, the user memory access happens without calling \texttt{nmi\_access\_okay}, which could be dangerous on x86 when it is using a different \texttt{mm} than the target.
Second, the implementation of \texttt{copy\_from\_user\_nofault} may try to acquire a spin lock under the configuration \texttt{CONFIG\_HARDENED\_USERCOPY}.
If invoked under NMI, this may cause a deadlock.

Triggering the vulnerability requires actually executing the eBPF program under a specific context.
Since Syzkaller is not efficient in generating eBPF programs with this helper and triggering them (on average, only 18 programs with this helper are executed by Syzkaller in a 48 hour as opposed to 6409 by \sname), it cannot find the vulnerability.

\begin{figure}
\vspace{0.04in}
\scriptsize
\begin{lstlisting}
       CPU0                    CPU1
       ----                    ----
lock(&htab->buckets[i].lock);
                               local_irq_disable();
                               lock(&rq->__lock);
                               lock(&htab->buckets[i].lock);
<Interrupt>
  lock(&rq->__lock);
\end{lstlisting}
\vspace{-0.1in}
\caption{\em lockdep warning showing how \CVEthree could cause a deadlock.}
\label{fig:cve_2023_0160}
\vspace{-0.15in}
\end{figure}

\def\tk#1{\textit{t\textsubscript{#1}}}
\def\lk#1{\textit{l\textsubscript{#1}}}

\noindent\textbf{\CVEthree} is a locking problem in the \texttt{BPF\_MAP\_TYPE\_SOCKHASH} map and helper functions for manipulating the map.
\texttt{BPF\_MAP\_TYPE\_SOCKHASH} is a type of hash map provided by eBPF for storing reference to \texttt{struct sock}.
It protects a bucket in a hash map from concurrent access using locks.
The map may be modified using both syscalls from user space programs and helper functions in eBPF programs in contexts that may be interrupted (i.e., the lock is IRQ-unsafe).
When an eBPF program that uses this type of map is invoked by an interrupt that already holds a lock, a deadlock could occur.

Figure~\ref{fig:cve_2023_0160} shows a potential deadlock scenario illustrated by \texttt{lockdep}, the runtime locking correctness validator in the Linux kernel.
First, an interruptible task on CPU0, \tk{1}, acquires the lock to a bucket, \lk{1}, when trying to modify the map.
Meanwhile, an interrupt on CPU1, \tk{2}, that holds another lock, \lk{2}, may be waiting to acquire \lk{1}.
Then, if \tk{1} is interrupted by an interrupt same as \tk{2} that also tries to acquire \lk{2}, a deadlock will occur since \tk{1} is interrupted and can no longer release \lk{1}, and \tk{2} will not be able to release \lk{2} as it is still waiting for \lk{1}.

\noindent\textbf{Vulnerability 4} is another locking rule violation in the \texttt{BPF\_MAP\_TYPE\_QUEUE} map and its corresponding helper functions similar to \CVEthree that could lead to deadlock.
The map provides FIFO storage for eBPF programs and the implementation also uses a lock to protect it from concurrent access.
Therefore, in a scenario similar to \CVEthree, a deadlock will occur.
lockdep also catches another way it can violate the locking rule.
Since the lock can be acquired in interrupt context or in context with interrupt enabled,
a task holding the lock can be interrupted by a interrupt that tries to acquire the same lock, resulting in a deadlock.

\section{Discussions on Future Maintenance of \Sname}
\label{sec:discussions}

In \sname, we build the semantic-correct program generation/mutation logic by studying the verifier as well as automatically extracting information (e.g., helper function definitions) needed by the logic.
Therefore, as the eBPF subsystem evolves, a question we ask ourselves is how \sname will adapt to changes in the eBPF subsystem, and more specifically, whether the process will require hefty manual efforts.
Fortunately, studying the verifier should be mostly a one-time effort given the safety guarantee provided by the verifier should remain mostly the same.
As the eBPF subsystem evolves, however, we can expect more program types, helper functions and maps will be added.
At that time, only the definition of the newly added program type, helper or map need to be added to \sname in order to support fuzzing them, which can be achieved automatically.

\section{Related Work} 
\label{sec:related}

\noindent\textbf{Fuzzing the eBPF subsystem.}~~
There are a couple of solutions for fuzzing the eBPF subsystem~\cite{BPFFuzzer, BPFFuzzer2, nilsen2020fuzzing, scannell2020ebpffuzzer}.
BPF fuzzer~\cite{BPFFuzzer} aims to test the eBPF verifier by leveraging LLVM fuzzer and sanitizer available in the user space.
To do so, it compiles the verifier in the user space and let LLVM fuzzer perform mutation-based coverage-guided fuzzing.
That is, a set of initial inputs need to be fed to the fuzzer, and new inputs will generated by mutating the initial inputs or corpus.
Sample eBPF programs from the Linux kernel source tree are used.
Overall, it manages to find one bug.

\cite{nilsen2020fuzzing} is a syntax-aware fuzzer targeting the eBPF subsystem in the Linux kernel based on Angora~\cite{chen2018angora}, a mutation-based fuzzer that requires a set of initial inputs.
Similar to ~\cite{BPFFuzzer}, it also uses the sample eBPF programs in the Linux kernel source tree.
It tracks interesting bytes that trigger new execution paths and then use gradient descent to guide the mutation.
However, it only found bugs in the \texttt{libbpf} in the user space.

\cite{scannell2020ebpffuzzer} is a bytecode-level semantic-aware fuzzer targeting the JIT compiler.
It works by generating bytecode with some awareness about the semantics imposed on the register states.
The generated eBPF program will first be checked by the verifier compiled in the user space using the approach in~\cite{BPFFuzzer}.
If the program is deemed valid by the verifier, it will be loaded into the kernel and JITed to see if it can trigger bugs.
The experiments shows only 0.77\% of generated eBPF programs are valid due to the limited semantics awareness, and the fuzzer manages to find one bug in the JIT compiler.

Another BPF fuzzer by Crump~\cite{BPFFuzzer2} is a syntax-aware differential fuzzer.
By feeding the generated eBPF programs to two different execution environment (i.e., native execution of JITed code and executed by the interpreter) and comparing the results, BPF fuzzer can detect bugs in the runtime if the outputs are different.
By using BPF fuzzer on RBPF, a user space BPF runtime implemented in Rust, it found two vulnerabilities.

Compare to previous work, \sname is a semantic-aware and dependency-aware generator-based fuzzer that is able to generate eBPF programs that pass the verifier efficiently.
Combining with efforts to attach and trigger the eBPF programs, \sname is able to reach deeply into the runtime and discover 4 new vulnerabilities.
To the best of our knowledge, \sname is the first work that cover all major eBPF runtime components.

\noindent\textbf{Formal verification of the eBPF subsystem.}~~
In addition to fuzzing, formal verification is another approach used to improve the safety of eBPF subsystem by verifying the correctness of components in the eBPF subsystem or implementing the components with proven correctness~\cite{gershuni2019simple, wang2014jitk, nelson2020specification, bhat2022formal, vishwanathan2021semantics}.

Prevail~\cite{gershuni2019simple} is an effort in implementing an alternative eBPF verifier in the user space with greater precision, which is adopted by eBPF for Windows~\cite{eBPFforWindows}.
JitK~\cite{wang2014jitk} implements an interpreter for cBPF with proven correctness.
JitSynth~\cite{van2020synthesizing} develops a tool that synthesizes eBPF bytecode into verified native RISC-V instructions.
Jitterbug\cite{nelson2020specification} models the eBPF JIT compiler in Rosette~\cite{torlak2013growing} and then uses it to verify the correctness of JIT compiler implementation of different architectures in the Linux kernel.
\cite{bhat2022formal} and \cite{vishwanathan2021semantics} try to verify the range analysis mechanism (i.e., \texttt{tnum}) using formal verification methods.
Interestingly, \CVEone discovered by \sname is due to a flaw in \texttt{tnum}.

We believe that fuzzing the eBPF runtime is orthogonal to the verification of eBPF.
Even in the future, when the verifier, JIT compiler and the interpreter can be implemented with proved correctness, we believe the rest of the runtime components are less likely to be verified due to their diverse functionality and increasing number.
Therefore, testing the runtime components should be a complementary effort in making the eBPF subsystem safer.

\noindent\textbf{Fuzzing language processors.}~~
Many efforts have been invested in improving fuzzing language processors~\cite{yang2011finding, holler2012fuzzing, veggalam2016ifuzzer, chen2018angora, aschermann2019nautilus, chen2021one, padhye2019semantic, park2020fuzzing, srivastava2021gramatron}, software that translate source code in high-level languages into lower-level languages.
Examples are compilers, JIT runtimes, and interpreters.
As mentioned earlier, eBPF is also a language processor.

These approaches can be categorized into mutation-based and generator-based.
For mutation-based fuzzers, initial inputs are required to generate new fuzzing inputs, which often suggests that the fuzzers have limited knowledge about the syntax and semantics.
For generator-based fuzzers, with some knowledge about the syntax or semantics, they are able to generate new inputs by themselves.

CSmith~\cite{yang2011finding} is a C compiler fuzzer that performs differential testing.
C code is randomly generated according to the grammar and then fed to different compilers.
After compilation and execution, the results are compared to determine if there are bugs in the compilers.
LangFuzz~\cite{holler2012fuzzing} is a mutation-based blackbox fuzzer that generates and mutates code fragments in the fragment pool, which is constructed by parsing codebases and test suites using a language-specific parser.
IFuzzer~\cite{veggalam2016ifuzzer} is a mutation-based fuzzer that try to generate new inputs using genetic programming.
Angora~\cite{chen2018angora} aims to improve the branch coverage by introducing several techniques that facilitate constraint solving without using symbolic execution.
The techniques are context-sensitive branch coverage, scalable byte-level taint tracking, search based on gradient descent, type and shape inference, and input length exploration.
Nautilus~\cite{aschermann2019nautilus} is a grammar-based fuzzer that does not rely on corpora.
It combines coverage-guided feedback to achieve higher fuzzing performance.
PolyGlot~\cite{chen2021one} improves the generic applicability of a grammar-aware fuzzer by performing mutation and analyses in IR level.
It first generates an IR translator using the BNF grammar of the language.
Then, inputs selected from corpus are lifted using the IR translator.
Constraints mutation will produce syntax-correct inputs and a semantic validator will further fix semantic errors.
Zest~\cite{padhye2019semantic} facilitates fuzzing of the semantic-stage of language processors by preserving the validity of syntax of inputs and using the validity feedback as guidance in addition to code coverage.
\cite{park2020fuzzing} is a JavaScript fuzzer that introduces aspect-preserving mutation that avoids destroying the semantic of corpus.
By stochastically preserving the structures and types in input corpus, it has better chance in generating inputs that are syntax-correct and semantic-correct.
Gramatron~\cite{srivastava2021gramatron} improves the efficiency of grammar-aware fuzzers by addressing two shortcomings of traditional parse tree-base fuzzer.
It uses grammar automaton to avoid biased sampling and aggressive mutation to avoid small-scale mutation.

\sname adopts the generator-based approach with the guidance of code coverage.
Besides, due to the complex eBPF semantics imposed by the verifier due to security reasons, we extract semantic rules from the verifier to make \sname grammar-aware, so that the fuzzing inputs are able to reach eBPF runtime efficiently.
We believe that these techniques for fuzzing language processors are orthogonal to our work and can be incorporated to \sname in the future to further improve the fuzzing effectiveness.

\section{Conclusions}
\label{sec:conclusions}

This paper introduced the BPF Runtime Fuzzer (\sname), a fuzzer that can satisfy the semantics and dependencies required by the verifier and the eBPF subsystem.
We addressed three important challenges in \sname: (1) generating semantic-correct eBPF programs, (2) generating syscall dependencies of eBPF programs, and (3) generate syscalls to attach and trigger eBPF programs.
Our experiments show, in 48-hour fuzzing sessions, \sname can successfully execute 8$\times$ more eBPF programs compared to Syzkaller.
Moreover, eBPF programs generated by \sname are much more expressive than Syzkaller's.
As a result, \sname achieves 101\% higher code coverage.
Finally, \sname has so far managed to find 4 vulnerabilities (some of them have been assigned CVE numbers) in the eBPF runtime, proving its effectiveness.

\section*{Acknowledgments}

The work was supported in part by NSF Awards \#1763172 and \#1846230 as well as Google's 2020 Android Security and PrIvacy REsearch (ASPIRE) Award.

\bibliographystyle{plain}
\bibliography{ardalan}

\pagebreak

\end{document}